\documentstyle[prd,aps,epsf]{revtex}

\begin{document}

\twocolumn[\hsize\textwidth\columnwidth\hsize\csname
@twocolumnfalse\endcsname

\title{Gravitational energy, dS/CFT correspondence and cosmic no-hair}

\author{Tetsuya Shiromizu$^{(a)}$
Daisuke Ida$^{(a)}$ and
Takashi Torii$^{(a), (b)}$
}

\address{
$^{(a)}$Research Center for the Early Universe (RESCEU),
The University of Tokyo, Tokyo 113-0033, Japan}
\address{$^{(b)}$ Advanced Research Institute for Science and
Engineering,
Waseda University, Shinjuku-ku, Tokyo 169-8555, Japan}



\maketitle

\begin{abstract}
The gravitational energy is examined in 
asymptotically de Sitter space-times. The positivity will be 
shown for certain cases. The de Sitter/CFT(dS/CFT) 
correspondence recently proposed and 
cosmic no-hair conjecture are testified in the aspect of 
the gravitational energy. From the holographic renormalization 
group point of view, the two conjectures are deeply 
connected with each other.
\end{abstract}
\vskip2pc]

 \vskip1cm

 \section{Introduction}\label{I}

 The origin of de Sitter entropy is getting to be central
issue\cite{deSitter}:
 {}From some considerations based on the euclidean quantum gravity or
quantum
 field theory in curved space-times, we obtain the Hawking-Bekenstein
 formula for the de Sitter space-times\cite{Area}. The entropy is
 proportional to the area of the cosmological horizon. Remember our
 success on the black hole entropy which is described by the state
 counting of string\cite{Stro}. How can we explain the de Sitter entropy
 in string
 theory?
 Apart from an issue of the de Sitter entropy,
 the fundamental study about the de Sitter space-times is important because
  our universe
 has experienced the inflationary phase in  very early stage, and
 the recent observation supports the existence of the positive cosmological
 constant\cite{SN}.
 The deep understanding of de Sitter geometry will tell us the origin
 of the vacuum energy.

 One way to this end might be given by
 de Sitter/CFT(dS/CFT) correspondence recently proposed\cite{dSCFT} as a
 possible extension of
 AdS/CFT correspondence\cite{AdSCFT}.
 Then, the de Sitter entropy should be explained by euclidean  CFT one.
 If we do not  care details,
 dS/CFT correspondence is naively
 expected
 from formal correspondence to  AdS/CFT
 through double Wick rotations.
 This relies on the fact that the de Sitter metric
 %
 \begin{eqnarray}
 ds^2=-dt^2+e^{2 {\sqrt {\Lambda/6}}t}( d\chi^2+dx^2+dy^2+dz^2)
 \label{flat}
 \end{eqnarray}
 %
 is obtained through the double Wick rotations ($t \mapsto i\chi$,
 $\chi \mapsto i t$) of AdS:
 %
 \begin{eqnarray}
 ds^2=d\chi^2+e^{2{\sqrt {-\Lambda/6}} \chi }(-dt^2+dx^2+dy^2+dz^2).
 \end{eqnarray}
 %
 For example, the trace anomaly of CFT is easily
 reproduced\cite{Nojiri}. Conversely, this means that things learned
 in study of the inflationary scenario is useful for
 AdS/CFT\cite{AdSCFT}, brane-worlds\cite{Lisa} and
 holographic renormalization group\cite{RG}.

 In this paper,
 we study the gravitational energy in asymptotically de Sitter
 space-time, and discuss its significance in the context
 of dS/CFT correspondence and the cosmic no-hair conjecture.
 The gravitational energy has been
 investigated so far in the inflationary universe. As Ashtekar and
 Das pointed out in AdS/CFT context\cite{Das}, it is
 natural to expect that the energy of euclidean CFT is related to the
 gravitational energy measured on the boundary of the de Sitter
 space-times.
 The gravitational energy will also be relevant for the stability of
 the de Sitter
 space-times
 and the corresponding euclidean CFT.
 As related topics, there is so called cosmic no-hair
 conjecture\cite{Area,Moss}.
 It is widely believed that
 initial inhomogeneity in
 our universe is rapidly stretched out and precisely evolve to
 the de Sitter space-times
 during the inflation.
 We can discuss the dynamics of
 such an evolution in terms of the gravitational energy\cite{AD2}.

 The rest of the present paper is organized as follows. In the
 Sec.~\ref{II}, we
 consider the asymptotically de Sitter
 space-times and their asymptotic behaviors. As an example we examine
 the $n$-dimensional Schwarzschild-de Sitter space-time.
 In the Sec.~III,
 we discuss the relation of the energy defined by Weyl tensor and
 the Abbott-Deser(AD) energy\cite{AD}. The AD energy has, however, a
 conceptual problem in asymptotically de Sitter space-times due to the
  non-existence
 of global static Killing vector.
 We propose
 the conformal energy associated with conformal Killing vector
 or spinor in Sec.~IV. We show
 the positive energy theorem for the conformal energy and
 give a simple relation to the AD energy.
 In Sec.~V,
 we discuss dS/CFT correspondence and cosmic no-hair conjecture as
 applications.
 Finally, we summarize our study in the
 Sec.~VI.

 \section{Asymptotically de Sitter space-times}\label{II}

 \subsection{Conformal infinity}

 In this section, we briefly give a review of the asymptotically
 de Sitter space-times following Ref. \cite{Tetsuya}.

 {\it Definition}: An $n$-dimensional space-time $(M, g)$
 will be said to be {\it asymptotically de Sitter} if there exists
 a manifold with boundary, $\hat M$, with the metric $\hat g$ such that
 (i) there exists a function $\Omega$ on $M$ such that $\hat g=\Omega^2 g$
 on $M$, (ii) ${\cal I}=\partial \hat M$ and $\Omega=0$ on ${\cal I}$,
 (iii) $g$ satisfies $n$-dimensional Einstein equation
 \begin{equation}
 R_{\mu\nu}-(1/2)g_{\mu\nu}R+\Lambda g_{\mu\nu}=8\pi T_{\mu\nu},
 \label{einstein}
 \end{equation}
 where
 $\Omega^{-(n-1)}{T_\mu}^\nu$ admits a smooth limit to ${\cal I}$ and
 $\Lambda>0$ is the positive cosmological constant.

 Then, we can show that
 %
 \begin{eqnarray}
 \hat g^{\mu\nu}\partial_\mu\Omega\partial_\nu\Omega
   & = & -\frac{2}{(n-1)(n-2)}\Lambda+
 \frac{\Omega^2}{n(n-1)}\hat R +O(\Omega^3) \nonumber \\
 &=:& -H^2+\frac{\Omega^2}{n(n-1)}\hat R+ O(\Omega^3)
 \label{n2}
 \end{eqnarray}
 %
 holds,
 where
 a hatted tensor field is regarded as a tensor field on $\hat M$.
 The extrinsic curvature $K_{\mu\nu}$ of the $\Omega={\rm constant}$
 hypersurfaces in $(M,g)$
 has the following behavior near the conformal infinity; for the trace
part
 %
 \begin{eqnarray}
 K=(n-1)H-\frac{n+1}{2n(n-1)H}\hat R \Omega^2 +O(\Omega^3),
 \end{eqnarray}
 %
 and for the trace-free part
 %
 \begin{eqnarray}
 {\sigma^\mu}_\nu={K^\mu}_\nu-\frac{1}{n-1}{q^\mu}_\nu K= O(\Omega^2),
 \end{eqnarray}
 where $q_{\mu\nu}$ denotes the metric of the $\Omega={\rm constant}$
 hypersurface.
 The derivation of these asymptotics is similar to
 that in the asymptotically AdS space-time ($\Lambda<0$),
 which is described in
 Ref. \cite{Ida}.

 \subsection{Spatial infinity and constant mean curvature
slices}\label{III}
 In asymptotically flat space-times, there is a natural concept of the
total
 gravitational
 energy (ADM energy \cite{ADM}) defined at the spatial infinity $i^0$.
 While in asymptotically de Sitter space-times, the conformal infinity
 ${\cal I}$
 consists of space-like
 hypersurface as seen from Eq.~(\ref{n2}), so that there are many spatial
 infinities on
 ${\cal I}$.
 We shall specify this by considering a flat chart of the
 de Sitter space-time as a reference
 background.
 It is useful because the intrinsic geometry of
 each constant time hypersurface of asymptotically de Sitter
 space-time will look like $(n-1)$-dimensional
 flat space. In order to see this in more detail, it is better to
 look at the Hamiltonian constraint on a space-like hypersurface $\Sigma$:
 %
 \begin{eqnarray}
 {}^{(n-1)}R+K^2-K_{\mu\nu}K^{\mu\nu}&=&16\pi T_{\mu\nu}n^\mu
n^\nu\nonumber\\
 &&{}+(n-1)(n-2)H^2,
 \label{hc}
 \end{eqnarray}
 %
 where $n^\mu$ is the future pointing unit normal vector to $\Sigma$.
 If there is a constant mean curvature slice
 $K=(n-1)H$ in an asymptotically de Sitter space-time, Eq. (\ref{hc})
becomes
 %
 \begin{eqnarray}
 {}^{(n-1)}R-\sigma_{\mu\nu} \sigma^{\mu\nu}=16\pi T_{\mu\nu}n^\mu n^\nu,
 \label{hccmc}
 \end{eqnarray}
 %
 which
 is exactly the same form  as the Hamiltonian constraint
 on a maximal ($K=0$) hypersurface  in asymptotically flat space-times.

 This observation indicates that the asymptotically de Sitter initial data
 can be formulated in a similar manner to the asymptotically flat case.
 More precisely, we shall call an initial data set $(\Sigma,q_{ij},K_{ij})$
 for the Einstein equation (\ref{einstein})
 {\em the asymptotically de Sitter initial data}, if it satisfies
 %
 \begin{eqnarray}
 q_{ij} = \Bigl(1+\frac{2m}{r^{n-3}} \Bigr)\delta_{ij}
 +O\Bigl(\frac{1}{r^{n-2}}  \Bigr)
 \end{eqnarray}
 %
 and
 %
 \begin{eqnarray}
 K_{ij}=Hq_{ij}+O\Bigl( \frac{1}{r^{n-2}} \Bigr),
 \end{eqnarray}
 %
 where
 $r=\sqrt {x^ix^i}$. Let ``$i^0$'' be spatial infinity at $r=\infty$. 
``$i^0$'' is presented by a point of the conformal infinity in the 
Penrose diagram. 
When we evaluate the total {\it finite} energy later, we must impose
 a stronger condition as follows:
 %
 \begin{eqnarray}
 K_{ij}=Hq_{ij}+O\Bigl( \frac{1}{r^{n-1}} \Bigr).
 \end{eqnarray}
 %

 As $K=0$ slices in asymptotically
 flat space-times\cite{Bartnik}, it is likely that we can prove the
 existence of $K=(n-1)H$ slices. In particular, it has been numerically
 confirmed that the 4-dimensional Schwarzschild-de Sitter is foliated by
 $K=3H$
 slices\cite{Nakao}. Moreover we can foliate the $n$-dimensional
 Schwarzschild-de Sitter space-times by $K_{\mu\nu}=Hq_{\mu\nu}$ slices
 (See Ref. \cite{BH} for 4-dimensional case):
 %
 \begin{eqnarray}
 ds^2  &=&
 -\left[\frac{1-m/2(ar)^{n-3}}{1+m/2(ar)^{n-3}} \right]^2
 d t^2
 \nonumber \\
 & &
 +a^2\left[ 1+\frac{m}{2(ar)^{n-3}} \right]^{4/(n-3)}
 \delta_{ij}dx^idx^j,\label{SdS}
 \end{eqnarray}
 %
 where $a=e^{Ht}$.
 This is a good example in the pedagogical point of view.
 Let us consider where the above coordinate covers. Best we can do is
 to find the coordinate transformation from the above chart to
 the static chart in which the metric is
 %
 \begin{eqnarray}
 ds^2=-F(R)dT^2+\frac{1}{F(R)}dR^2+R^2 d\Omega_{n-2}^2,
 \end{eqnarray}
 %
 where $F(R)=1-2m/R^{n-3}-H^2R^2$. The corresponding coordinate
 transformation is given by
 %
 \begin{eqnarray}
 R=ar \left[1+\frac{m}{2(ar)^{n-3}} \right]^{2/(n-3)},
 \end{eqnarray}
 %
 and
 %
 \begin{eqnarray}
 T & = & t+H \int^R \frac{R}{F(R)}\left(1-\frac{2m}{R^{n-3}}
\right)^{-1/2}dR.
 \end{eqnarray}
 %
 Figure 1 shows the Penrose diagram.

 \section{Abbott-Deser energy and Weyl tensor}

 In this section, we show that the conserved Abbott-Deser(AD)
energy\cite{AD}
 is identical to that defined by Weyl tensor on the constant mean curvature
 slices. The latter will play
 a central role in dS/CFT issue later.
 The AD energy has the expression \cite{AD2,AD,Tess}
 %
 \begin{eqnarray}
 E_{\rm AD}=M_{\rm ADM}+\Delta P_{\rm ADM}(\bar \xi ), \label{AD}
 \end{eqnarray}
 %
 %
 \begin{eqnarray}
 \Delta P_{\rm ADM}(\bar \xi) & =& P_{\rm ADM}(\bar \xi)-
 \bar P_{\rm ADM}(\bar \xi) \nonumber \\
 & = & \frac{1}{8\pi}\oint_{S^{n-2}_\infty} d\bar S_i
 ({\pi^i}_j-{{\bar{\pi}}^i}_{\;j} ) \bar \xi^j,
 \end{eqnarray}
 %
 where $\pi_{ij}=K_{ij}-Kq_{ij}$, $\bar\pi_{ij}=-(n-2)Hq_{ij}$
 is that defined on the background,
 $\bar \xi^i$ is the space component of the static Killing vector
 of the background de Sitter space-time in the flat chart. $S_\infty^{n-2}$ 
denotes the $n-2$-sphere at the spatial infinity ``$i^0$''. 
 (See the Appendix A for the definition of the AD energy.)

 On the other hand, we
 may expect that the gravitational energy is measured
 by the tidal force (the electric part of the Weyl tensor).
 It is natural to expect a direct relation between the AD energy and the
Weyl
  tensor.
 In fact, this has been confirmed for the four-dimensional spherically
 symmetric case
 \cite{AD2}. We will extend this to general cases.

 The total gravitational energy associated with a slice $\Sigma$
 is defined in terms of the
 $n$-dimensional Weyl tensor ${}^{(n)}C_{\mu\alpha\nu\beta}$ by
 %
 \begin{eqnarray}
 E_W & := & -\frac{1}{8\pi} \oint_{S_\infty^{n-2}}ar
 {}^{(n)}C_{\mu\alpha\nu\beta}\hat r^\mu \hat r^\nu n^\alpha n^\beta
 dS^{n-2} \nonumber \\
 &= & M_{\rm ADM}
 -\frac{(n-3)H}{8\pi} \oint_{S^{n-2}_\infty}
 ar \Bigl(k_{\mu\nu} \nonumber \\
 & & +\frac{1}{n-3}q_{\mu\nu}k \Bigr) \hat r^\mu \hat r^\nu dS^{n-2},
 \label{physical}
 \end{eqnarray}
 %
 where, $S^{n-2}_\infty$ is regarded as the $(n-2)$-sphere
 at the spatial infinity ($i^0$) of $\Sigma$,
 $\hat r^\mu$ is the unit outward normal vector to
 $S^{n-2}_\infty$ (specified by the
 condition $\hat r^\mu n_\mu=0$)
 and $k_{\mu\nu}:=K_{\mu_\nu}-Hq_{\mu\nu}$.
 Here we used 
 %
 \begin{eqnarray}
 M_{\rm ADM} & = & -\frac{1}{8\pi} \oint_{S_\infty^{n-2}} ar
 {}^{(n-1)}R_{\mu\nu}
 \hat r^\mu \hat r^\nu dS^{n-2}\nonumber \\
 & = & \frac{1}{16\pi} \oint_{S^{n-2}_\infty}
 (\partial^i q_{ij}-\partial_j {q^i}_i)dS^j,
 \end{eqnarray}
 %
 and
 %
 \begin{eqnarray}
 E_{\mu\nu}:&=&{}^{(n)}C_{\mu\alpha\nu\beta}n^\alpha n^\beta
 \nonumber \\
 &=&{}^{(n-1)}R_{\mu\nu}+(n-3)H \Bigl(k_{\mu\nu}+\frac{1}{n-3}
 q_{\mu\nu}k \Bigr)
 \nonumber \\
 && +kk_{\mu\nu}-{k_\mu}^\alpha k_{\alpha\nu}.
 \end{eqnarray}
 %
Those can be derived following the argument of Ref. \cite{Anne}. 
If $K=(n-1)H$ holds on $\Sigma$, we have $k=0$ so that
Eq.~(\ref{physical})
 becomes
 %
 \begin{eqnarray}
 E_W=M_{\rm ADM}-\frac{n-3}{8\pi}\oint_{S_\infty^{n-2}} (Har) k_{\mu\nu}\hat
 r^\mu
 \hat r^\nu dS^{n-2}.
 \end{eqnarray}
 %
 In the same way, we obtain the same expression for the AD energy.
 Here we have required $k_{\mu\nu}=O(1/r^{n-1})$
 to make the second term
 finite.\footnote{The corresponding term in
 asymptotically flat space-times is $\int k_{\mu\nu} \hat r^\mu \hat r^\nu
 d^{n-2}S$ which is automatically finite because of the
 absence of the factor, $Har$.}
 This fall off might be faster than
 naively expected
 (See Sec.~IIB).
 However, on
 $K=(n-1)H$ slices,
 we
 have a simple expression: $E_W=E_{\rm AD}=M_{\rm ADM}$.

 Here we have one serious and well known problem. AD energy defined
 here is associated
 with the static Killing vector. The Killing vector is spacelike outside of
 the cosmological horizon and evaluate the total energy outside
 the cosmological horizon. This is not congenial to the
 term ``energy'', because the energy must be measured by the
 {\it timelike} observers. In the next section, we discuss a new
 interpretation of the energy which supports the use of the AD energy
 in asymptotically de Sitter space-times.

 \section{Energy and conformal Killing vector}

 In this section we introduced the new energy associated
 with the conformal Killing vector/spinor to overcome the
 conceptual problem of the AD energy.

 \subsection{Conformal Killing vector and spinor}

 The de Sitter space-time has also a conformal
 static
 Killing vector, which is everywhere timelike, deduced from its conformal
 flatness.
 In terms of the conformal time the de Sitter metric can be written as
 %
 \begin{eqnarray}
 ds^2=a^2(\tau) [-d\tau^2+\delta_{ij}dx^i dx^j],
 \end{eqnarray}
 %
 and the conformal static Killing vector is\footnote{There is an ambiguity 
related to the trivial rescaling freedom. But, the freedom is renormalised 
into the definition of $a$.}
 %
 \begin{eqnarray}
 \xi=\partial_\tau=a \partial_t
 \end{eqnarray}
 %
 satisfying conformal Killing equation
 \begin{equation}
 \mbox \pounds_\xi g_{\mu\nu} =2Hg_{\mu\nu}.
 \end{equation}
 Correspondingly, the conformal
 Killing spinor $\epsilon$ is defined by
 %
 \begin{eqnarray}
 \nabla_\mu \gamma_\nu \epsilon +\nabla_\nu \gamma_\mu \epsilon=
 \frac{1}{2}g_{\mu\nu} \nabla_\alpha (\gamma^\alpha \epsilon ). \label{CKS}
 \end{eqnarray}
 %
 We can easily check that $\bar \epsilon \gamma^\mu \epsilon$ is
 the conformal Killing vector.
 Equation~(\ref{CKS}) is invariant under the conformal transformation
 $\tilde g_{\mu\nu}=\Omega^2 g_{\mu\nu}$ with $\tilde \epsilon =
 \Omega^{1/2}\epsilon$.

 Let us consider the pure de Sitter case.
 We take $\Omega=a^{-1}$
 and then $\tilde \epsilon = a^{-1/2}\epsilon_0$, where $\epsilon_0$
 is a constant Killing spinor satisfying $\gamma^{\hat 0}\epsilon_0=i
 \epsilon_0$
 (${\hat 0}$ denotes the time-component with respect to the orthonormal
 basis).
 The associated vector is $\xi_0{}^\mu=\bar \epsilon_0 \gamma^\mu
 \epsilon_0=(\partial_\tau)^\mu$.

 It is helpful to
 see the relation between $\tilde \nabla_i \tilde \epsilon $ and
 $\nabla_i \tilde \epsilon$:
 %
 \begin{eqnarray}
 \tilde \nabla_i \tilde \epsilon=\Bigl(\nabla_i+\frac{1}{2}
 \frac{\dot \Omega}{\Omega}\gamma_i \gamma^{\hat 0}  \Bigr)\tilde \epsilon,
 \end{eqnarray}
 %
 where we set $\Omega = a^{-1}=e^{-Ht}$.
 The second term is expected to compensate the
 cosmological term in $\nabla_i$, because the conformally transformed
 space-time will be asymptotically flat.

 Finally, we look at the features of the asymptotically flatness for the
 conformally transformed space-times in  detail. As an example,
 we take the $n$-dimensional Schwarzschild-de Sitter space-time given by
 Eq.~(\ref{SdS}).
 For $ar \gg m $ the conformally transformed metric becomes
 %
 \begin{eqnarray}
 \tilde g_{\mu\nu}dx^\mu dx^\nu & \simeq &
 -\left[1-\frac{2m}{(ar)^{n-3}} \right]d\tau^2 \nonumber \\
 & & +\left[1+\frac{2m}{(n-3)(ar)^{n-3}} \right]
 \delta_{ij}dx^i dx^j.
 \end{eqnarray}
 %
 The point, which is different from the standard asymptotically flat
 space-times,
 is the scale factor dependence appearing together with the radial
 coordinate $r$ like $m/[a(t)r]^{n-3}$.

 \subsection{Conformal energy}

 Let us consider the total gravitational
 energy in conformally transformed
 space-times.

 First, we propose the energy associated with the {\it conformal Killing
 spinor}. Since we are considering the asymptotically flat space-times,
 we might be able to prove the positive energy theorem (See Ref.
\cite{Tess}
 for asymptotically dS space-times.) if all the asymptotic arguments are
 correct.
 The conformal energy is defined by
 %
 \begin{eqnarray}
 {\tilde E}_W :& = &
 \int d\tilde S_{\mu\nu}( \tilde \epsilon^\dagger \tilde
 \gamma^{\mu\nu\alpha} \tilde \nabla_\alpha \tilde \epsilon
 +{\rm h.c.} ) \nonumber \\
 & = & \int d\tilde
 \Sigma\Bigl[ |\tilde \nabla_i \tilde \epsilon |^2+ 4\pi
 \tilde T_{\hat 0 \hat 0}|\tilde \epsilon|^2+4\pi \tilde T_{\hat 0 \hat i}
 {\tilde {\bar \epsilon}} \gamma^{\hat 0} \gamma^{\hat i}\tilde
 \epsilon \Bigr],
 \end{eqnarray}
 %
 where
 %
 \begin{eqnarray}
 8\pi \tilde T_{00}=8\pi T_{00}+(n-2)H\left[(n-1)H-K\right] 
 \label{T00},
 \end{eqnarray}
 %
 and
 %
 \begin{eqnarray}
 \tilde T_{0i}=T_{0i}.
 \end{eqnarray}
 %
 For a $K=(n-1)H$ slice we get $T_{00}=\tilde T_{00}$ from Eq.~(\ref{T00}).
 Thus, ${\tilde E}_W$ is manifestly non-negative on the $K=(n-1)H$ slices if
 the dominant energy condition for the stress-energy tensor $T_{\mu\nu}$ is
 assumed.
 In addition, we can prove that the
 physical space-time is the de Sitter space-time
 with $\Lambda=(n-1)(n-2)H^2/2$ if ${\tilde E}_W=0$ is satisfied.

 As usual $\tilde E_W$ is the ADM energy plus ADM momentum for the
conformally
 transformed space-time. This means that $\tilde E_W$ is written by
 the electric part of the Weyl tensor\cite{Anne}\footnote{The 
conformally transformed spacetimes is asymptotically flat spacetimes 
in our sense. This means that we can use the argument in asymptotically 
flat spacetimes.}:
 %
 \begin{eqnarray}
 {\tilde E}_W = -\frac{1}{8\pi}\oint_{\tilde S^{n-2}_\infty}r \tilde
 C_{\mu\alpha\nu\beta} \tilde n^\mu \tilde n^\nu
 {\tilde {\hat r^\alpha}}{\tilde {\hat r^\beta}} d\tilde S^{n-2}.
 \end{eqnarray}
 %
 Since the Weyl tensor is invariant for the conformal transformation,
 we can show ${\tilde E}_W =a^{3-n}E_W$, where $E_W$ is defined by
 Eq.~(\ref{physical}) and identical to AD energy on $K=(n-1)H$ slices
 together with the condition $k_{\mu\nu}=O(1/r^{n-1})$. Hence we
 could prove the positivity of AD energy.\footnote{At
 first glance this seems to contradict with two examples with
 the negative AD energy given in Ref.~\cite{AD2}.
 Here, to remove any confusions, we insist that there is no
 contradiction. There are two distinctions between the present
 study and paper \cite{AD2}. In Ref.~\cite{AD2} the authors did not
 pay attention very much on the finiteness of AD energy. The spherical
 example with the negative AD energy in Ref.~\cite{AD2} does not
 have the constant mean curvature slices. This means that the
 local energy condition for $\tilde T_{\mu\nu}$ does not hold.

 As a result, we obtained the following corollary:
 {\it there are no $K=(n-1)H$ slices satisfying the
 strong fall-off condition such that the AD energy is negative}}.

 \section{Applications}

 Based on the Abbott-Deser energy, we carefully examined
 the gravitational energy in asymptotically de Sitter
 space-times so far. In this section as examples we will use the
 energy to testify dS/CFT correspondence and cosmic
 no-hair conjecture.

 \subsection{dS/CFT correspondence}

 Here we 
 shall consider the dS/CFT correspondence in terms of
 the gravitational energy. 

 The stress-energy tensor of CFT can be evaluated as \cite{CFT}
 %
 \begin{eqnarray}
 \langle T_{\mu\nu}\rangle_{\rm CFT} & = & (n-2)Hq_{\mu\nu}+K_{\mu\nu}-
 q_{\mu\nu}K
 \nonumber \\
 & & +\frac{1}{(n-3)H}{}^{(n-1)}G_{\mu\nu}+\cdots. \label{CFT2}
 \end{eqnarray}
 %
 The derivation is quite similar to the quasi-local energy
 proposed by Brown and York\cite{York}, who used the Hamilton-Jacobi
 formalism,
 so that Eq.~(\ref{CFT2}) has also the concept of the quasi-local energy.
 Hence we expect that it expresses the total gravitational energy at 
the conformal infinity. If this expectation is correct,
 it will be a demonstration of dS/CFT correspondence.

 Physically the total energy is measured
 via the tidal force at the infinity,
 which is encoded in the electric part of the Weyl tensor:
 %
 \begin{eqnarray}
 E_{\mu\nu}
 & = & {}^{(n-1)}G_{\mu\nu}+(n-3)H(k_{\mu\nu}-q_{\mu\nu}k)
 \nonumber \\
 & & +kk_{\mu\nu}-k_{\mu\alpha}{k_\nu}^\alpha
 +\frac{1}{2}q_{\mu\nu}(k_{\alpha\beta}k^{\alpha\beta}-k^2 ).
 \label{electric}
 \end{eqnarray}
 %
 The total gravitational energy
 can be evaluated with $E_{\mu\nu}$ \cite{Anne}.
 As seen in Sec.~III
 the extrinsic curvature $K_{\mu\nu}$ has the asymptotic behavior
 %
 \begin{eqnarray}
 K_{\mu\nu}=H q_{\mu\nu}+k_{\mu\nu}
 \end{eqnarray}
 %
 where
 %
 \begin{eqnarray}
 k_{\mu\nu}=O\left(\frac{1}{r^{n-1}}\right)
 \end{eqnarray}
 %
 near the conformal infinity.

 Accordingly, the stress-energy tensor of CFT is written as
 %
 \begin{equation}
 \langle T_{\mu\nu} \rangle_{\rm CFT}
 =\frac{1}{(n-3)H}{}^{(n-1)}G_{\mu\nu}+(k_{\mu\nu}-q_{\mu\nu}k)
 +\cdots. \label{CFT}
 \end{equation}
 %
 {}From Eqs.~(\ref{electric}) and (\ref{CFT}), we have
 %
 \begin{eqnarray}
 (n-3)H \langle T_{\mu\nu} \rangle_{\rm CFT}
 -E_{\mu\nu}&=&
 -kk_{\mu\nu}+\cdots\nonumber\\
 &=&O(1/r^{2n-2})
 \end{eqnarray}
 %
 near the conformal infinity.

 Integrating over sphere $S^{n-2}$ at the spatial infinity ``$i^0$''
 introduced in the previous section, we obtain
 %
 \begin{eqnarray}
 & & \oint_{S_\infty^{n-2}} dS^{n-2}r \Bigl[(n-3)H \langle
 T_{\mu\nu} \rangle_{\rm CFT}
 -E_{\mu\nu}\Bigr] \hat r^\mu \hat r^\nu \nonumber \\
 & & =\oint_{S_\infty^{n-2}} dS^{n-2}r (-kk_{\mu\nu}+\cdots)
 \hat r^\mu \hat r^\nu =O\Bigl(\frac{1}{r^{n-2}} \Bigr),
 \end{eqnarray}
 or
 \begin{equation}
 E_W=(n-3)H\oint_{S_\infty^{n-2}} dS^{n-2}r \langle T_{\mu\nu}
 \rangle_{\rm CFT}\hat r^\mu
 \hat r^\nu.
 \end{equation}
 Hence the energy of CFT is identical with the total gravitational energy.
 This can be contrasted to
 Ashtekar and Das's claim\cite{Das} for AdS/CFT correspondence.

 \subsection{Cosmic no-hair}

 Let us remember that $\tilde M$ is positive definite and
 ${\tilde E}_W=E_{\rm AD}/a^{n-3}$.
 Since
 $E_{\rm AD}$ is conserved, we have $\tilde M \to 0$
 as $a \to \infty$ for $ n>3$.
 This shows the cosmic no-hair
 property, since  $\tilde M=0$ implies that
 the physical space-time is the de Sitter space-time.
 As a result, inhomogeneities
 on $K=(n-1)H$ slices
 will be stretched
 as time passes long enough. In other words, the geometry near the 
conformal infinity looks like the deSitter spacetime. 

 The cosmic no-hair is
 closely related to dS/CFT correspondence.
 In the same way as AdS/CFT, the method of holographic renormalization
 group for the euclidean CFT is applied,
 where the time coordinate $t$ of the de Sitter metric is
 related to the renormalization scale.
 Hence it is important to show
 that the space-times with the positive cosmological constant
 evolves toward geometry like $ds^2=-dt^2+e^{2Ht}q_{ij}({\bf x})dx^i
 dx^j $ which is just the statement of the
 cosmic no-hair conjecture.

 \section{Summary}

 In this paper, we examined Abbott-Deser energy.
 First we showed that AD energy is identical with the energy defined by
 the electric
 part of the Weyl tensor. Since the electric part expresses the
 tidal force, this is physically desirable result.

 Next we introduced the new energy associated with the conformal
 Killing vector to overcome one serious problem of AD energy.
 Since AD energy refers to the static Killing vector of
 the de Sitter space-time,
 which is spacelike outside the
 cosmological horizon, so that the physical meaning of AD energy outside
 the cosmological horizon is not clear.
 The point is that de Sitter
 space-time has the global static conformal Killing vector
 since it is conformal to the flat space-time.
 We therefore expect that an asymptotically de Sitter space-time is
 conformal to some asymptotically flat space-time, and that
 corresponding asymptotically conformal static Killing vector gives
 a natural definition of the gravitational energy.
 As an example, we consider the conformal energy in terms of the Nester
 formula,
 which has some nice properties;
 On a $K=(n-1)H$ slice,
 the conformal energy is positive definite and vanishes iff the
 physical space-time is the de Sitter space-time.
 Furthermore, with the strong fall off condition on the traceless part of
 the extrinsic curvature $k_{\mu\nu}$,
 the conformal energy agrees with the AD energy up to the scale factor.

 We
 discussed the role of gravitational energy in dS/CFT correspondence
 and cosmic no-hair conjecture.
 We showed that the CFT energy is
 same as the gravitational energy.
 This provides us one evidence
 for dS/CFT. The cosmic no-hair was discussed by using the
 energy defined in conformally transformed space-times.
 The conformal energy becomes zero near the future infinities
 due to the scale factor dependence of the energy,
 and then from the positive energy argument, we can conclude
 that 
 the physical space-time approaches the de Sitter space-time.
 This feature supports the cosmic no-hair conjecture.

 Our arguments relies on the existence of
 a constant mean curvature slice of $K=(n-1)H$
 and the fall-off condition for the extrinsic curvature
 $k_{\mu\nu}$.
 We expect the existence of such a constant mean curvature slice for a
 wide class
 of space-times \cite{henkel}.
 $K=(n-1)H$ slices are also suitable for the set-up of  dS/CFT
 correspondence.
 The fall off condition on the traceless part
 of the extrinsic curvature might be rather strong,
 though it is essential to make the energy finite.
 Under this condition,
 the total
 gravitational energy is well-defined and dS/CFT
 correspondence  works well.

 \section*{Acknowledgements}

 We would like to thank K. Nakao, H. Ochiai and  Y. Shimizu for their
 discussion. TS is grateful to G. W. Gibbons for his useful
 suggestion in 1998.
 TS's work is partially supported by Yamada Science Foundation.

 \appendix

 \section{Abbott-Deser energy}

 We decompose the metric into $n$-dimensional de Sitter metric
 $\bar g_{\mu\nu}$ and the rest $h_{\mu\nu}$; $g_{\mu\nu}=\bar g_{\mu\nu}
 + h_{\mu\nu}$. The basics to define the energy is that the Einstein
equation
 is written as $R^{\mu\nu}_L-\frac{1}{2}\bar g^{\mu\nu}R_L-\Lambda
h^{\mu\nu}
 =(-\bar g)^{-1/2}{\cal T}^{\mu\nu}$,
 where $h^{\mu\nu}=\bar g^{\mu\alpha} g^{\nu\beta} h_{\alpha\beta}$,
 $R^{\mu\nu}_L$ is the linear part of Ricci tensor $R_{\mu\nu}$
 with respect to $h_{\mu\nu}$. From the Bianchi identity, we see
 $\bar \nabla_\mu {\cal T}^{\mu\nu}=0$ and then $\partial_\mu
 ({\cal T}^{\mu\nu}\bar \xi_\nu )=0$, where $\bar \xi^\mu$ is the
 static Killing vector of the background de Sitter space-time.
 Thus we can define the conserved energy as
 %
 \begin{eqnarray}
 E_{\rm AD}=\frac{1}{8\pi} \int d^{n-1}x{\cal T}_{0\mu} \bar \xi^\mu.
 \end{eqnarray}
 %

 If the background geometry is given by the de Sitter metric
 $\bar g= -dt^2+a(t)^2 \delta_{ij} dx^i dx^j$, where
 $a(t)=e^{Ht}$, the background
 Killing vector has the component $\bar \xi^\mu
 =(1, -H x^i)$.
 Then, Eq.~(\ref{AD}) gives the refined form of the AD energy.
 For the $n$-dimensional Schwarzschild-de Sitter space-time
 Eq.~(\ref{SdS}), the AD energy is just the mass parameter:
 %
 \begin{eqnarray}
 E_{\rm AD}=\frac{(n-2)\pi^{(n-3)/2}}{4\Gamma\Bigl( \frac{n-1}{2} \Bigr)}m.
 \end{eqnarray}
 %

\section{Conformal Komar mass}

We consider the stationary case such as Kerr-de
 Sitter space-time.
 In this case, we can consider
 the {\em conformal Komar mass} defined below. (See the original paper
 \cite{Komar} for the
 Komar mass.)
 Let us consider the conformal transformation
 $\tilde g_{\mu\nu}=a^{-2}g_{\mu\nu}$ with $a=e^{Ht}$. The stationary
 Killing vector $\xi=\partial_t$ of the original space-time
 becomes conformal Killing vector of the conformally transformed space-time
 satisfying
 %
 \begin{eqnarray}
 \mbox \pounds_\xi \tilde g_{\mu\nu}=2H\tilde g_{\mu\nu}.
 \end{eqnarray}
 %
 Then, we can show that
 %
 \begin{eqnarray}
 \tilde \nabla_\mu \tilde \nabla_\nu \xi_\alpha=-\tilde
R_{\nu\alpha\mu\beta}
 \xi^\beta,
 \end{eqnarray}
 %
 and
 %
\begin{equation}
 \tilde \epsilon^{\rho\sigma\alpha_1 ... \alpha_{n-2}}
 \tilde \nabla_\sigma ( \tilde \epsilon_{\alpha_1...\alpha_{n-2}\mu\nu}
 \tilde \nabla^\mu
 \xi^\nu) 
 =-2(n-2)! \tilde {R^\rho}_\sigma \xi^\sigma
 \end{equation}
 %
 hold.
 Here 
 the relation between $\tilde R_{\mu\nu}$ and $R_{\mu\nu}$ is given by
 %
 \begin{eqnarray}
 \tilde R_{00}=R_{00}+KH,
 \end{eqnarray}
 %
 and
 %
 \begin{eqnarray}
 \tilde R_{0i}=R_{0i}.
 \end{eqnarray}
 %
 Thus $\tilde R_{\mu\nu}$ does not contain the leading term from the
 positive cosmological constant term if we consider
 $K=(n-1)H$ slices, namely
 \begin{equation}
 \tilde R_{00}=8\pi \left(T_{00}-\frac{T}{n-2}g_{00}\right).
 \end{equation}
 Moreover, in the vacuum region, we see
 %
 \begin{eqnarray}
 \tilde \epsilon^{\rho\sigma\alpha_1...\alpha_{n-2}}
 \tilde \nabla_\sigma ( \tilde \epsilon_{\alpha_1...\alpha_{n-2}\mu\nu}
 \tilde \nabla^\mu
 \xi^\nu)=0.
 \end{eqnarray}
 %
 Hence we can define the conformal Komar mass by
 %
 \begin{eqnarray}
 \tilde M & = &  -\frac{1}{8\pi } \oint_{\tilde S} \tilde
 \epsilon_{\alpha_1 ...\alpha_{n-2} \mu\nu} \tilde
 \nabla^\mu \xi^\nu d\tilde S^{\alpha_1...\alpha_{n-2}}
 \nonumber \\
 & = & \frac{1}{4\pi} \int_{\tilde \Sigma}\tilde R_{\mu\nu}\tilde t^\mu
 \xi^\nu d\tilde V \nonumber \\
 & = & 2\int_{\tilde \Sigma}
 \Bigl(T_{\mu\nu}-\frac{1}{n-2}g_{\mu\nu}T-\Lambda g_{\mu\nu} \Bigr)
 \tilde t^\mu n^\nu d \tilde V, \label{Komarint}
 \end{eqnarray}
 %
 which gives a conserved energy.
 {}From the integrand in the last line in Eq. (\ref{Komarint}), we can
 read that the vacuum energy is automatically subtracted.

 \section{Stress tensor of CFT}

 There are several way to derive Eq. (\ref{CFT2}). We adopt the
 path integral procedure\cite{Emanuel}:
 %
 \begin{eqnarray}
 Z & = & \int {\cal D}g e^{iS_g} \nonumber \\
   & = & \int {\cal D}q \int_{g|_B=q} {\cal D}g e^{iS_g} \nonumber \\
   & = & \int {\cal D}q e^{i\Gamma_{\rm CFT}+iS_{\rm ct}(q)},
 \end{eqnarray}
 %
 where $\Gamma_{\rm CFT}$ is the effective action of CFT living on the
 boundary. $S_{\rm ct}$ is the counter term introduced to make the
 action finite. It is determined using the Hamilton-Jacobi formalism.

\newpage

\begin{figure}[t]
\label{figure1}
\vspace*{-5mm}
\begin{center}
\epsfxsize=2.5in
~\epsffile{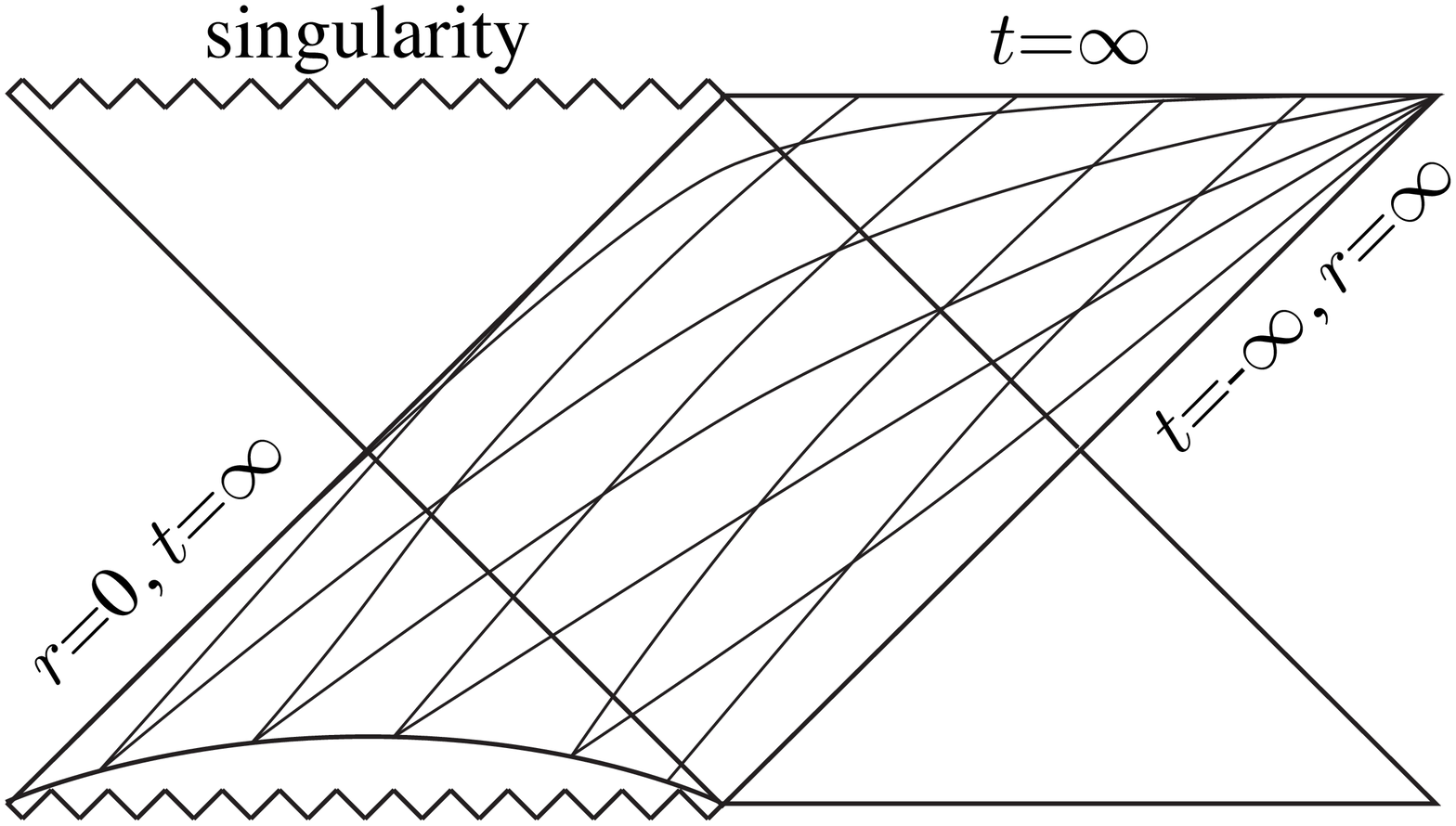}
\end{center}
\caption{This is the Penrose diagram of the Schwarzschild-de Sitter
space-times. The $K_{\mu\nu}=Hq_{\mu\nu}$ slice foliates a
part of the whole space-times. $t=const.$ and $r=const.$ lines
are drawn schematically.}
\end{figure}

 \end{document}